\documentclass[a4paper,aps,prl,twocolumn,showpacs,a4paper]{revtex4}

\usepackage{epsf,subfigure}
\usepackage{amsmath,graphicx}
\usepackage[T1]{fontenc}
\usepackage[latin1]{inputenc}
\usepackage{times,ulem,color,bm}
\usepackage[squaren,Gray]{SIunits}
\usepackage{pdfpages}

\newcommand{\cora}[1]{\textcolor{black}{#1}}
\newcommand{\corb}[1]{\textcolor{black}{#1}}

\usepackage[pdfstartview=FitV,bookmarks=true,
            linktocpage=true,colorlinks=true,
            urlcolor=blue,linkcolor=blue,
            citecolor=blue]{hyperref}

\begin{document}

\title{Full transmission and reflection of waves propagating through a maze of disorder}
%Open and closed scattering channels through an elastic disordered waveguide

\author{Benoît Gérardin}
\author{Jérôme Laurent}
\author{Arnaud Derode}
\author{Claire Prada}
\author{Alexandre Aubry} \email{alexandre.aubry@espci.fr}
\affiliation{ESPCI ParisTech, PSL Research University, CNRS, Univ Paris Diderot, Sorbonne Paris Cité, Institut Langevin, UMR 7587, 1 rue Jussieu, F-75005 Paris, France}
\date{\today}

\begin{abstract}
Thirty years ago, theorists showed that a properly designed combination of incident waves could be fully transmitted through (or reflected by) a disordered medium, based on the existence of propagation channels which are essentially either closed or open (\textit{bimodal} law). In this Letter, we study elastic waves in a disordered waveguide and present a {direct} experimental evidence of the bimodal law. Full transmission and reflection are achieved. The wave-field is monitored by laser interferometry and highlights the interference effects that take place within the scattering medium.

\end{abstract}

%\pacs{42.25.-p,07.60.-j,78.67.-n,42.15.Fr}

\maketitle

Light travelling through thick clouds, electrons conducting through metals or seismic waves in the earth crust are all examples of waves propagating through disordered materials. Energy transport by waves undergoing strong scattering is usually well described by diffusion theory. However, this classical picture neglects interference effects that may resist the influence of disorder. Interference is responsible for fascinating phenomena in mesoscopic physics. On the one hand, it can slow down and eventually stop the diffusion process, giving rise to Anderson localization~\citep{anderson,Lagendijk}. On the other hand, it can also help waves to find a way through a maze of disorder~\cite{pendry}. Actually, a properly designed combination of incident waves can be completely transmitted through a strongly scattering medium, as suggested by Dorokhov and others more than twenty years ago~\citep{dorokhov,imry,pendry2,beenakker}. This prediction has recently received a great deal of attention mostly due to the emergence of wave-front shaping techniques in optics~\cite{Mosk_review}. 

In order to address the open channels (\textit{i.e.}, to achieve full energy transmission) across a disordered wave guide, one has to perform a complete measurement of the scattering matrix $\mathbf{S}$. The $\mathbf{S}$-matrix relates the input and output of the medium~\cite{beenakker}.
%A wave incident on the disordered region is described in the basis of the scattering channels by a vector of coefficients
%However, the wavefields can be tamed in order to take advantage of the complexity of propagation media to for instance focus waves. This is realized in the temporal domain using the concept of the time reversal mirror, which was originally demonstrated in acoustics and then in electromagnetism, or in the spatial domain using the wavefront shaping techniques developed in optics. The coherent control of the incident wave-field allows for instance to make multiple scattering paths interfere constructively at any location behind a strongly scattering sample. 
% 
It fully describes wave propagation across a scattering medium. It can be generally divided into blocks containing transmission and reflection matrices, $\mathbf{t}$ and $\mathbf{r}$, with a certain number $N$ of input and output channels. Initially, random matrix theory (RMT) has been successfully applied to the transport of electrons through chaotic systems and disordered wires~\cite{beenakker}. However, the confrontation between theory and experiment has remained quite restrictive since specific input electron states cannot be addressed in practice. %Hence one cannot have access to the $\mathbf{S}$-matrix experimentally. 
On the contrary, a coherent control of the incident wave-field is possible in classical wave physics. 
%This has been done for ages in acoustics with arrays of transducers or in electromagnetism with arrays of antennas. More recently, the emergence of wave front shaping techniques in optics has allowed the spatio-temporal control of light across disordered samples. 
Several works have demontrated the ability of measuring the \textbf{S}-matrix, or at least some of its subspaces, in disordered media, whether it be in acoutics~\citep{sprik,aubry2009random,aubry_2014}, electromagnetism~\cite{shi,davy} or optics~\citep{popoff2010measuring,choi,popoff2,choi2}.

The existence of open channels has been revealed by investigating the eigenvalues $T$ of the Hermitian matrix $\mathbf{tt^{\dag}}$. Theoretically, their distribution should follow a bimodal law~\citep{dorokhov,imry,beenakker}, exhibiting two peaks. The highest one, around $T \sim 0$, correspond to closed (\textit{i.e.} strongly reflected) eigenchannels. At the other end of the spectrum ($T \sim 1$), there are $g$ open eigenchannels. $g = Nl^*/L$ is the dimensionless conductance~\cite{feng2}, $L$ is the sample thickness, and $l^*$ is the transport mean free path. By exciting selectively open or closed channels, a nearly complete transmission~\cite{choi3,rotter}, reflection or absorption~\cite{chong} of waves can be achieved. It means that a designed wave-front can be fully transmitted or, on the contrary, fully reflected by a scattering medium, which is in total contradiction with the classical diffusion picture. {Although some indirect evidence of bimodality have been pointed out experimentally as \textit{e.g.}, the reduction of the shot noise power in electrical conductors~\citep{beenaker2,henny}}, these remarkable interference effects have never been directly observed so far. Indeed, the bimodal distribution relies on the conservation of energy (\textit{i.e.} $\mathbf{S}$ is a unitary matrix). In other words, all the channels should be addressed at the input and measured at the output~\cite{goetschy}. In optical experiments, the finite numerical aperture of the illumination and detection systems limits the angular coverage of the input and output channels~\cite{popoff_2013}. In acoustics or electromagnetism, the spatial sampling of measurements has not been sufficient to have access to the full $\mathbf{S}-$matrix so far~\citep{sprik,aubry2009random,shi}.
%Moreover, a wave guide geometry is required in order to avoid energy leaks from the sides of the scattering sample. For mainly all these reasons, the bimodal distribution has never been measured. Instead %

In this Letter, we present experimental measurements of the full $\mathbf{S}$-matrix across a disordered elastic wave guide. To that aim, laser-ultrasonic techniques have been used in order to obtain a satisfying spatial sampling of the field at the input and output of the scattering medium. The unitarity of the $\mathbf{S}-$matrix is investigated and the eigenvalues of the transmission matrix are shown to follow the expected bimodal distribution. Moreover, full experimental transmission and reflection of waves propagating through disorder are achieved. The wave-fields associated to the open and closed channels are monitored within the scattering medium by laser interferometry to highlight the interference effects operating in each case.

\begin{figure}[ht]
\centering
\includegraphics[width=\columnwidth]{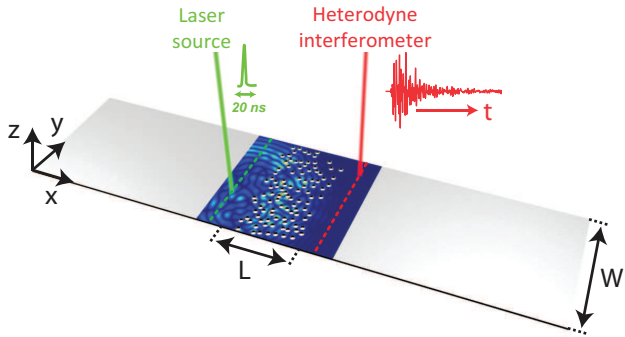}
\caption{Experimental set up. The $\mathbf{S}-$matrix is measured in the time-domain, between two arrays of points placed 5 mm away from each side of the disordered slab. The array pitch is $0.8$ mm ($\sim \lambda/4$). Flexural waves are generated on each point by a pulsed laser \textit{via} thermo-elastic conversion over a focal spot of 1 mm$^2$. The normal component of the plate vibration is measured with an interferometric optical probe. The laser source and the probe are both mounted on 2D translation stages.}% Once the measurement is done, the $\mathbf{S}-$matrix is decomposed in the Fourier domain and in the basis of the plate modes according the procedure described in the Method.}
\label{fig1}
\end{figure}

\cora{We study here the propagation of \cora{elastic} waves across a duralumin plate (aluminium alloy) of dimension $500\times40\times0.5$ mm$^3$ (see Fig.~\ref{fig1}). The homogeneous plate is a waveguide in which randomness is introduced by drilling circular holes with diameter 1.5 mm and concentration 11 cm$^{-2}$, distributed over a thickness $L=20$ mm \cora{and the whole width of the plate $W=40$ mm}. The $\mathbf{S}$-matrix associated to the disordered slab is measured with the laser-ultrasonic set-up described in Fig.~\ref{fig1}. Elastic waves are generated in the thermoelastic regime \cite{Krishnaswami} by a pumped diode Nd:YAG laser providing pulses having a \unit{20}{\nano\second} duration and \unit{2.5}{\milli\joule} of energy. The out-of-plane component of the local vibration of the plate is measured with a heterodyne interferometer \cite{Krishnaswami}. In the frequency range of interest (0.32 to 0.37 MHz, $\Delta f= 0.05$ MHz), the plate thickness ($d=0.5$ mm) is small compared to the wavelength ($\lambda~\sim~$3.5 mm).}

\cora{Three types of elastic waves can propagate through the plate: shear horizontal (SH), extensional and flexural waves \cite{royer}. For flexural waves, polarization is perpendicular to the plane of the plate whereas it is in-plane for SH and extensional waves. As a consequence, only flexural waves are measured by the heterodyne interferometer. Moreover, as scatterers consist in through holes, there cannot be conversion of SH/extensional modes into flexural modes \textit{via} scattering for symmetry reasons~\cite{diligent}. Hence, the experimental setup shown in Fig.~\ref{fig1} allows a measurement of the $\mathbf{S}$-matrix associated to the flexural modes in the plate.}

{The first step of the experiment consists in measuring the impulse responses between two arrays of points placed on the left and right sides of the disordered slab. \corb{Each impulse response is averaged over 128 laser shots in order to reduce additive electronic noise \cite{supp}}. The array pitch is {$0.8$ mm} (\textit{i.e.} $< \lambda/2$) which guarantees a satisfying spatial sampling of the wave field. \corb{The impulse responses between any two points of the same array form the time-dependent reflection matrices, $\mathbf{r}$ and $\mathbf {r'}$, from left to left and right to right, respectively. The set of impulse responses between the two arrays yield the time-dependent transmission matrices $\mathbf{t}$ (from left to right) and $\mathbf{t'}$ (from right to left)}. From these four matrices, one can build the $\mathbf{S}-$matrix in a \textit{point-to-point} basis,}
\begin{equation}
\mathbf{S}= \left ( \begin{array}{cc}
\mathbf{r} & \mathbf{t'} \\
\mathbf{t} & \mathbf{r'}
\end{array} 
 \right )
 \label{S_blocks}
\end{equation}
\cora{A temporal Fourier transform of $\mathbf{S}$ is then performed over a time range $\Delta t=120$ $\mu$s that excludes the echoes due to reflections on the ends of the plate. Frequency components that are spaced by more than the correlation frequency $\delta f$ give rise to uncorrelated speckle patterns. In our disordered sample, we have measured $\delta f \sim $ 0.017  MHz. Hence, the number $N_f=\Delta f / \delta f$ of independent scattering matrices $\mathbf{S}$ that are obtained over the frequency bandwidth is 3. The correlation frequency also yields an estimation for the Thouless time ($\tau_D \sim 1/ \delta f \sim 60$ $\mu$s) \textit{i.e.}, the mean time it takes for a wave to cross the sample through its zigzag motion. We check that $\tau_D \sim \Delta t /2 $ and that most of the energy has escaped from the sample when the measurement is stopped. The next step of the experimental procedure consists in decomposing the $\mathbf{S}$-matrices in the basis of the flexural modes of the homogeneous plate. These eigenmodes and their wavenumbers have been determined theoretically using the thin elastic plate theory~\citep{supp,cross,santamore}. They are normalized so that each of them carries unit energy flux across the plate section. The transformation of the $\mathbf{S}-$matrix from the point-to-point basis to the channel basis is described in details in the supplemental material \cite{supp}.} Fig.~\ref{fig2a} displays an example of matrix $\mathbf{S}$ in the channel basis recorded at frequency $f=0.36$ MHz. Despite its overall random appearance, one can see the residual ballistic wave-front that slightly emerges along the diagonal of the transmission matrices. The matrix $\mathbf{S}$ also exhibits long-range correlations that will account for the bimodal behavior of the transmission/reflection matrices.\\

\begin{figure}[!ht]
\centering
\subfigure{\includegraphics[width=\columnwidth]{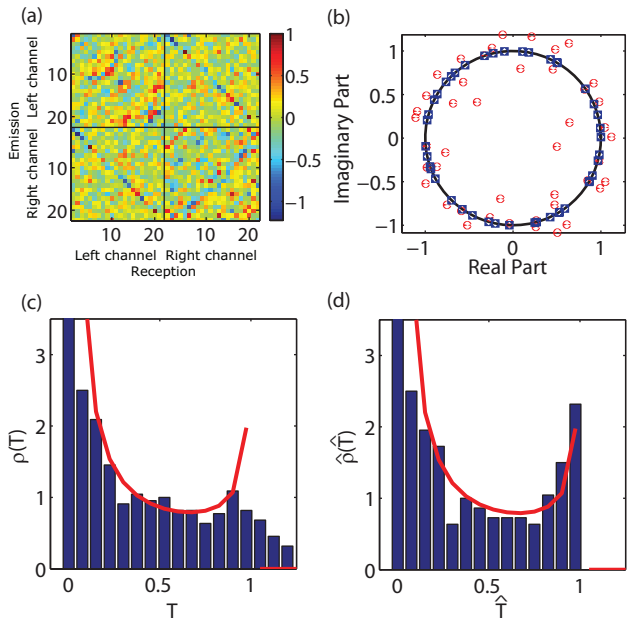}\label{fig2a}}
\subfigure{\label{fig2b}}
\subfigure{\label{fig2c}}
\subfigure{\label{fig2d}}
\caption{\textbf{a}, Real part of the $\mathbf{S}$-matrix measured at $f=0.36$ MHz. The black lines delimit transmission and reflection matrices as depicted in Eq.~\ref{S_blocks}. \textbf{b}, Eigenvalues $s_i$ (red dots) and $\hat{s}_i$ (blue squares) of the measured and normalized scattering matrices, $\mathbf{S}$ and $\mathbf{\hat{S}}$, respectively. The eigenvalues are displayed in the complex plane. The black continuous line denotes the unit circle. \textbf{c}-\textbf{d}, Transmission eigenvalue histograms, $\rho(T)$ and $\hat{\rho}(\hat{T})$, averaged over the frequency bandwidth. Both distributions are compared to the bimodal law $\rho_b$ (red continuous line, Eq.~\ref{eq:bimodal}).}
\label{fig2}
\end{figure}

%\begin{figure}[htbp]
%\center
%\includegraphics[width=8cm]{Figures/fig3.eps}
%\caption{Bimodal law. Mettre tous les détails de la figure (bande de fréquence correspondante à chacune des figures. En fait je la changerais, je trouve que celle sur le pwt 'Fevrier 2014' est plus jolie et montre un meilleur fit avec la loi bimodale.}
%\label{fig3}
%\end{figure}

\begin{figure*}[!ht]
\centering
\begin{minipage}[c]{0.75\textwidth}
\subfigure{\includegraphics[width=\textwidth]{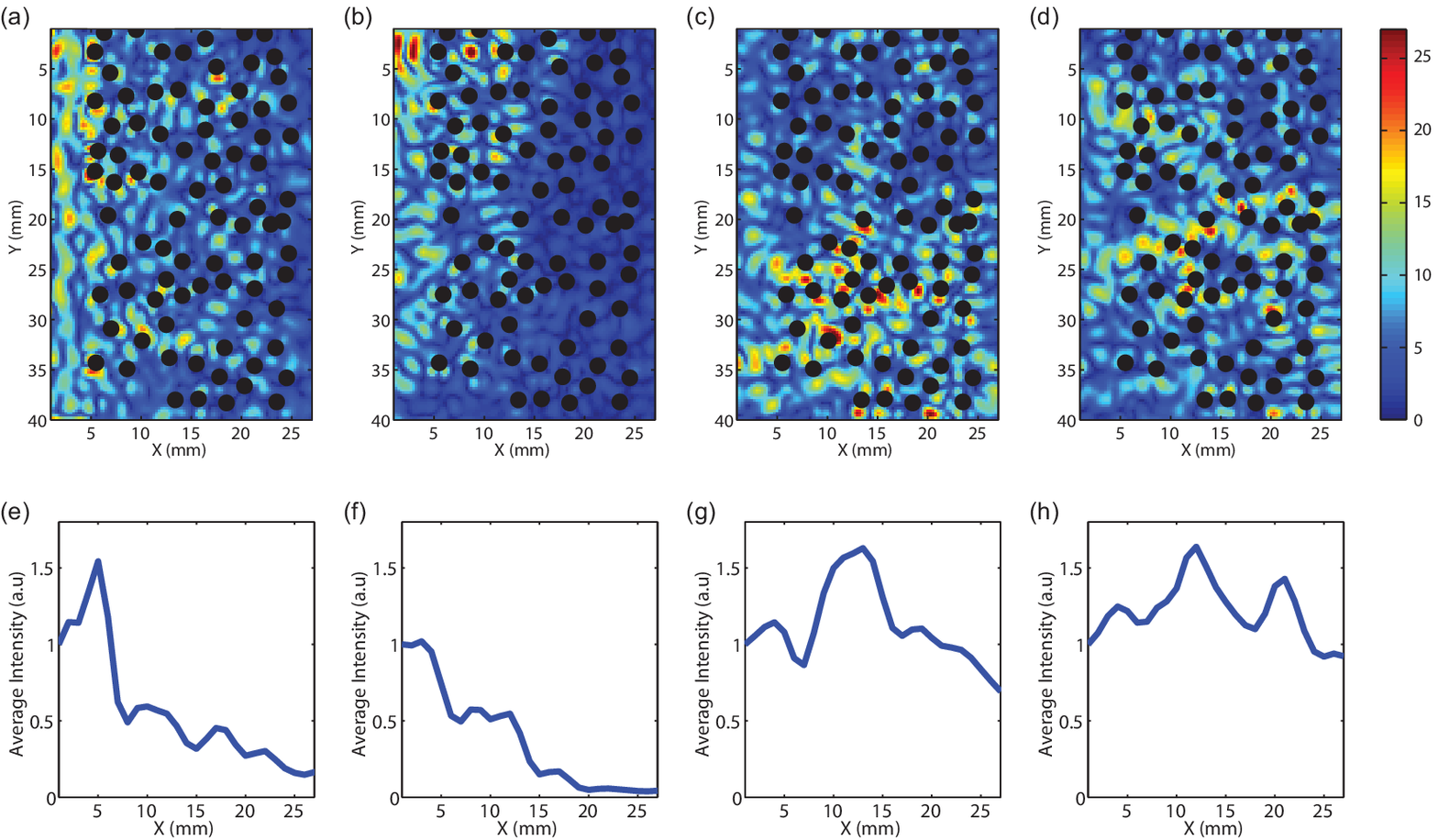}\label{fig3a}}
\subfigure{\label{fig3b}}\subfigure{\label{fig3c}}
\subfigure{\label{fig3d}}\subfigure{\label{fig3e}}
\subfigure{\label{fig3f}}\subfigure{\label{fig3g}}\subfigure{\label{fig3h}}
\end{minipage}\hfill
\begin{minipage}[c]{0.23\textwidth}
\caption{Absolute value of the wave-field in the scattering medium at $f=0.36$ MHz associated to \textbf{a}, an incident plane wave, \textbf{b}, a closed eigenchannel, \textbf{c}, an open eigenchannel deduced from the measured $\mathbf{S}$-matrix and \textbf{d}, an open eigenchannel deduced from the normalized $\mathbf{\hat{S}}$-matrix. The corresponding intensities averaged over the wave guide section (y-axis) are shown versus depth $x$ in lower panels \textbf{e}-\textbf{h}. They are all normalized by the intensity at the plane of sources ($x=0$).}
\label{fig3}
\end{minipage}
\end{figure*}
Theoretically, energy conservation would imply that $\mathbf{S}$ is unitary. In other words, its eigenvalues should be distributed along the unit circle in the complex plane. The eigenvalues $s_i$ of the $\mathbf{S}$-matrix at $f=0.36$ MHz are displayed in the complex plane in Fig.~\ref{fig2b}. The dispersion of these eigenvalues around the unit circle questions the validity of the energy conservation assumption. \corb{Yet duralumin is known for its weak absorption properties ($\sim 1.7$ dB.m$^{-1}$ in the 0.1-0.5 MHz frequency bandwidth \cite{Hsieh})}. Radiation losses through conversion of elastic waves into acoustic waves in surrounding air can also be neglected ($\sim 1$ dB.m$^{-1}$ at 0.36 MHz \cite{footnote2}). 
%\corb{Actually, the non-unitarity of $\mathbf{S}$ mainly comes from the laser-induced deformation of the plate surface during the time of the experiment. To obtain a satisfying signal-to-noise ratio, we actually work close to the ablation threshold \cite{Krishnaswami}. The slight change of the surface state after each laser shot leads to fluctuations in the thermo-elastic conversion efficiency. It manifests itself as a multiplicative noise in the measured $\mathbf{S}$-matrix}. 
\corb{Actually, the non-unitarity of $\mathbf{S}$ is due to the experimental noise. Its magnitude, its nature and its various origins are described in details in the Supplemental material \cite{supp}.}
In the following, we compensate for this undesirable effect by building a virtual scattering matrix $\mathbf{\hat{S}}$ with the same eigenspaces as $\mathbf{S}$ but with normalized eigenvalues (see Fig.~\ref{fig2b}), such that
\begin{equation}
\label{eq2}
\hat{s}_{i}=s_i/\left | s_i  \right |\mbox{, for }i=1, \cdots, N
\end{equation}
The phase information is thus unchanged and only the amplitude is normalized to meet the energy conservation requirement. 
%Note that the normalized scalar product between $\mathbf{S}$ and $\mathbf{\hat{S}}$ is of 0.95. 
We will show that this normalization based on energy conservation is essential to observe the open channels.

We first focus on the transmission eigenvalues $T$ and $\hat{T}$ computed from $\mathbf{S}$ and $\mathbf{\hat{S}}$, respectively. \cora{Their sum} directly provides an estimation of the dimensionless conductance of the disordered slab: $g=\sum_i T_i \sim 8$ and $\hat{g} = \sum_i \hat{T}_i \sim 8.1$. \cora{As we have access to $N = 2W/\lambda \sim 22$ independent channels, this yields a ratio $L/l^* \sim 2.75$ according to the Ohm's law}. The distributions of the transmission eigenvalues, $\rho(T)$ and $\hat{\rho}(\hat{T})$, are estimated by averaging their histograms over the frequency bandwidth. Figs.~\ref{fig2c}-\ref{fig2d} show the comparison between these distributions and the bimodal law $\rho_b$ which is theoretically expected in the diffusive regime~\cite{beenakker},
\begin{equation}
\label{eq:bimodal}
\rho_b(T)=\frac{g}{2T\sqrt{1-T}}
\end{equation}
{Strictly speaking, our system is not in a fully diffusive regime ($L \sim 2.75 l^*$). Yet,} the measured eigenvalue distribution $\rho(T)$ is in correct agreement with the bimodal law $\rho_b(T)$ with a large peak around $T=0$ associated to the closed channels and a smaller peak around $T=1$ associated to the open channels (Fig.~\ref{fig2c}). Nevertheless, some channels exhibit a transmission coefficient superior to 1, which violates energy conservation. This is explained by the non-unitarity of $\mathbf{S}$ pointed out in Fig.~\ref{fig2b}. On the contrary, after normalization of $\mathbf{S}$ (Eq.~\ref{eq2}), all transmission eigenvalues $\hat{T}$ are repelled below 1 and the eigenvalue distribution $\hat{\rho}(\hat{T})$ closely follows the expected bimodal law (Fig.~\ref{fig2d}). This confirms that the unitarity of $\mathbf{S}$ is decisive and that experimental noise can prevent from recovering the bimodal law experimentally. \cora{In the supplemental material \cite{supp}, a numerical simulation confirms that the normalization of $\mathbf{S}$ (Eq.~\ref{eq2}) allows to retrieve almost completely the open channels, thus cancelling the detrimental effect of noise. We will now prove it experimentally by probing the open eigenchannels deduced from $\mathbf{S}$ and $\mathbf{\hat{S}}$.} 
%Note also that our system is not strictly speaking in a fully diffusive regime ($L \sim 3 l_e$). Yet the experimental distribution of transmission eigenvalues follows the bimodal law, confirming that it also applies for thickness of few scatterring mean free paths. \\
%\begin{figure*}[!ht]
%\centering
%\subfigure{\includegraphics[width=\textwidth]{fig3.eps}\label{fig3a}}
%\subfigure{\label{fig3b}}\subfigure{\label{fig3c}}
%\subfigure{\label{fig3d}}\subfigure{\label{fig3e}}
%\subfigure{\label{fig3f}}\subfigure{\label{fig3g}}\subfigure{\label{fig3h}}
%\caption{Absolute value of the wave-field in the scattering medium at $f=0.36$ MHz associated to \textbf{a}, an incident plane wave, \textbf{b}, a closed eigenchannel, \textbf{c}, an open eigenchannel deduced from the measured $\mathbf{S}$-matrix and \textbf{d}, an open eigenchannel deduced from the normalized $\mathbf{\hat{S}}$-matrix. The corresponding intensities averaged over the wave guide section (y-axis) are shown versus depth $x$ in lower panels \textbf{e}-\textbf{h}. They are all normalized by the intensity at the plane of sources ($x=0$).}
%\label{fig3}
%\end{figure*}

Whereas the eigenvalues of $\mathbf{tt^{\dag}}$ yield the transmission coefficients of each eigenchannel, the corresponding eigenvector provide the combination of incident modes that allow to excite this specific channel. Hence, the wave-field associated to each eigenchannel can be measured by backpropagating the corresponding eigenvector. To that aim, the scattering medium is scanned with the interferometric optical probe~\cite{supp}. As a reference, the wave-field induced by an incident plane-wave is shown in Fig.~\ref{fig3a}. Fig.~\ref{fig3e} plots the corresponding intensity averaged along the plate section ($y-$axis) as a function of depth $x$. Fig.~\ref{fig3b} displays the wave-field associated to a closed eigenchannel ($T \sim 0$). The wave is fully reflected back to the left. This shows that the incoming wavefield has been successfully tailored to fit the randomness of the waveguide, making it a nearly perfectly reflecting medium. Beyond one transport mean-free path, the intensity decreases very rapidly in agreement with numerical simulations~\cite{choi}. Figs.~\ref{fig3c}-\ref{fig3d} display the propagation of open eigenchannels ($T \sim 1$ and $\hat{T} \sim 1$) deduced from the measured and normalized matrices $\mathbf{S}$ and $\mathbf{\hat{S}}$, respectively. Both wave-fields clearly show the contructive interferences that help the wave to find its way through the maze of disorder. The corresponding intensity profiles are shown in Figs.~\ref{fig3g}-\ref{fig3h}. In both cases, the field intensity increases inside the medium. For the eigenchannel derived from the $\mathbf{S}$-matrix, the intensity measured at the output of the scattering medium is smaller than at the input (Fig.~\ref{fig3g}). Experimental noise in the $\mathbf{S-}$matrix prevents from addressing a fully opened channel. On the contrary, Fig.~\ref{fig3h} illustrates how nicely the normalized $\mathbf{\hat{S}}$-matrix gives access to a fully open eigenchannel with equal intensities at the input and output of the scattering medium. The incoming wavefield has been successfully designed to make the scattering slab completely transparent.

%\begin{figure}[htbp]
%\center
%\includegraphics[width=8.5cm]{fig4.eps}
%\caption{\textbf{a}, The distribution $\rho(T)$, obtained for an array pitch $p=3 \lambda/4$, is compared to the bimodal law $\rho_b(T)$ (red continuous line, Eq.~\ref{eq:bimodal}). \textbf{b}, The distribution $\rho(T)$, obtained for $p=\lambda$, is compared to the Mar\u{c}enko-Pastur law~\cite{marcenko,tulino} (red continuous line).}
%\label{fig4}
%\end{figure}

%\noindent {A point we would like to discuss is the decisive role of disorder-induced spatial correlations of the wave-field in the emergence of open eigenchannels. Fig.~\ref{fig4} displays the distribution of transmission eigenvalues when the array pitch $p$ exceeds the fundamental limit $\lambda/2$. When $p=3\lambda/4$, Fig.~\ref{fig4}a indicates that the open eigenchannels cannot be fully addressed. And, for $p = \lambda$, the distribution tends towards the Mar\u{c}enko-Pastur law (Fig.~\ref{fig4}b) that applies for random matrices with uncorrelated entries~\citep{marcenko,tulino}. Hence, the access to open eigenchannels strongly depends on the ability of measuring properly the spatial correlations of the wave-field, which requires a spatial sampling below a half wavelength. In other words, a complete channel control is needed to retrieve the bimodal law, as recently pointed out by Goetschy and Stone~\cite{goetschy}.}

{From a more general point of view, we want to highlight that being able to address the open eigenchannels is the key to make the best of a scattering medium, for many applications. For instance in digital telecommunication, the maximum information that can be conveyed with no error is Shannon's capacity $C$. $C$ is actually determined by the number $g$ of open eigenchannels, in a multiple input-multiple output scheme~\citep{derode2,goetschy}. Hence a complete channel control allows to reach the maximum information transfer rate. An other example is wave focusing through a multiple scattering medium, whether it be by time-reversal~\citep{derode,FinkPhysTo97}, phase conjugation~\cite{yaqoob} or wavefront-shaping optimization as the one used by Vellekoop and Mosk~\cite{vellekoop2007focusing}. At one frequency, the maximum contrast between the focal spot and the background intensity at the output is given by the effective number $N_{eff}= 3g/2$ of channels contributing to the transmitted field~\citep{vellekoop2,davy}. Again, the maximum is reached when the channel control is complete and the bimodal law is retrieved. Hence, the phase conjugation process is fully optimized only if all the open eigenchannels are properly addressed.}\\

%To coclude, we want to stress on thet fact that this experimental study gives access for the first time to fully open and closed eigenchannels through a scattering medium. However one can wonder whether they could be addressed in more practical situations and in other fields of wave physics such as optics or electromagnetism. First, the energy conservation assumption implies weakly dissipative systems. From a theoretical point of view, a perfect knowledge of the modes on the left and right leads of the system is required in order to renormalize them propertly such each of them carry the same energy flux. From a practical point of view, a complete channel control is decisive~\cite{goetschy}. Fig. displays the transmission eigenvalue histogram $\rho(T)$ obtained when the spatial sampling of the field exceeds $\lambda/2$. Most of the disorder induced correlations are lost and the bimodal behavior disappears. The transmission eigenvalue distribution tends to approach the Marcenko-Pastur law which applies when the transmission matrix elements are independently distributed.  In optics, a complete channel control implies a complete angular coverage of the input and output wave-fields. It also implies a perfect control of the polarization at emission and reception. At last, a sufficient signal-to-noise ratio is needed since noise can be detrimental to the recover of fully open eigenchannels, as pointed out in this paper. 

{In summary,} this experimental study {allows the direct observation} of fully open and closed eigenchannels through a scattering medium. In agreement with theoretical predictions, transmission eigenvalues across a disordered system are shown to follow a bimodal law. The wave-field associated to each eigenchannel is measured and illustrates the remarkable interference mechanisms induced by disorder. From a fundamental point of view, such measurements will allow to check a whole set of RMT predictions that could not have been confronted to experiment so far. The transition towards Anderson localization should lead to an extinction of fully opened eigenchannels~\citep{dorokhov,imry,beenakker} and the occurence of necklace states~\cite{pendry}. From a more practical point of view, this work paves the way towards an optimized control of wave-fronts through scattering media in all fields of wave physics, whether it be for communication, imaging, focusing, absorbing, or lasing purposes.\\

The authors wish to thank M. Davy, G. Lerosey, S. Rotter, S. Popoff, and A. Goetschy for fruitful discussions, as well as A. Souilah for his technical help. The authors are grateful for funding provided by LABEX WIFI (Laboratory of Excellence within the French Program Investments for the Future, ANR-10-LABX-24 and ANR-10-IDEX-0001-02 PSL*). B.G. acknowledges financial support from the French ``Direction Générale de l'Armement''(DGA).
%
%\bibliography{biblio} %% database .bib file

\begin{thebibliography}{36}
\bibitem[]{footnote2}
\bibinfo{note}{Radiation losses have been estimated numerically with the DISPERSE software \cite{disperse}}
\end{thebibliography}

%\vspace{5mm}
%
%\includepdf[pages={1,2,3,4}]{supplementary_material.pdf}

\end{document}